\pdfoutput=1
\documentclass[aps,prl,twocolumn,superscriptaddress]{revtex4-2}

\usepackage{graphicx}
\usepackage{hyperref}

\begin{document}


\title{Chaotic Diffusion in Delay Systems: Giant Enhancement by Time Lag Modulation}


\author{Tony Albers}
\email[]{tony.albers@physik.tu-chemnitz.de}
\author{David M\"uller-Bender}
\email[]{david.mueller-bender@mailbox.org}
\author{Lukas Hille}
\affiliation{Institute of Physics, Chemnitz University of Technology, 09107 Chemnitz, Germany}
\author{G\"unter Radons}
\email[]{radons@physik.tu-chemnitz.de}
\affiliation{Institute of Physics, Chemnitz University of Technology, 09107 Chemnitz, Germany}
\affiliation{Institute of Mechatronics, 09126 Chemnitz, Germany}


\date{\today}

\begin{abstract}
We consider a typical class of systems with delayed nonlinearity, which we show to exhibit chaotic diffusion.
It is demonstrated that a periodic modulation of the time-lag can lead to an enhancement of the diffusion constant by several orders of magnitude.
This effect is the largest if the circle map defined by the modulation shows mode locking and more specifically, fulfills the conditions for laminar chaos.
Thus we establish for the first time a connection between Arnold tongue structures in parameter space and diffusive properties of a delay system.
Counterintuitively, the enhancement of diffusion is accompanied by a strong reduction of the effective dimensionality of the system.
\end{abstract}


\maketitle

Chaotic diffusion is a well-known deterministic phenomenon in nonlinear dynamical systems, where the state variable shows normal or anomalous diffusive motion.
Research in this field started decades ago for low-dimensional Hamiltonian systems \cite{chirikov1979,lichtenberg1992,zacherl1986,geisel1987}.
A fundamental understanding of the mechanisms leading to diffusion was achieved for systems with strong dissipation,
where the dynamics can be modeled by one-dimensional iterated maps \cite{geisel1982,schell1982,fujisaka1982,geisel1985}.
Recently, the inequivalence of ensemble and time averages
became a major topic in diffusion related research in deterministic systems and in statistical models of the latter \cite{bel2006,metzler2014,albers2014,akimoto2015,albers2018,meyer2018},
meaning that chaotic diffusion in low-dimensional systems is nowadays well understood.
The opposite is true for deterministic diffusion in systems with infinite-dimensional evolution equations:
Only recently, diffusion in non-random systems, where the dynamics is governed by partial differential equations, was observed and analyzed \cite{cisternas2016,cisternas2018,albers2019_1,albers2019_2}.
In this context, the diffusing objects are exploding solitons.
Because of the general relation between partial differential equations and time-delayed systems \cite{giacomelli1996,buenner1997,yanchuk2017,marino2018,marino2020},
such solitons can also be found in the latter \cite{garbin2015,marconi2015,brunner2018,semenov2018,schelte2019,yanchuk2019}, but no deterministic diffusion was observed so far.
For time-delay systems, chaotic diffusion was hardly investigated, except for a very specific system in \cite{wischert1994,schanz2003,sprott2007,dao2013_1,dao2013_2}
and recent work \cite{lei2011,mackey2021} on an integrated version of the Ikeda equation \cite{ikeda1980}.
This is surprising in view of the highly developed mathematical theory of time-delay systems \cite{hale1993,diekmann1995,hale2002}
and their broad applicability covering all branches of science \cite{kuang1993,schoell2007,erneux2009,lakshmanan2011} and engineering \cite{stepan1989,michiels2007,erneux2009}.

In this Letter, we will show that for a large class of delay systems with linear instantaneous and nonlinear delayed term deterministic chaotic diffusion is possible,
and that, for instance, a simple modulation of the delay can increase the diffusion constant by several orders of magnitude.
We study these unexpected phenomena for systems with scalar evolution equations of the form
\begin{equation}
\label{eq:DDE}
\frac{1}{\Theta}\frac{dx(t)}{dt}=-x(t)+f(x(R(t))),
\end{equation}
where $R(t)=t-\tau(t)$ is the retarded argument and $\tau(t)$ is a time-varying delay.
In the following, we use as a representative variation
\begin{equation}
\label{eq:retarded_argument}
R(t)=t-\tau_0-\frac{A}{2\pi}\sin(2\pi t),
\end{equation}
which for $A=0$ includes the case of constant delay $\tau_0$.
Well-known prototype systems are obtained for specific choices of $f$ in Eq.~(\ref{eq:DDE}), such as the Mackey-Glass equation, a model for blood cell production \cite{mackey1977},
or the Ikeda equation, describing the phase dynamics in an optical ring cavity \cite{ikeda1980}.
Various other nonlinearities $f(x)$ have been considered in the literature \cite{lakshmanan2011}.

While most investigations considered constant delay \cite{farmer1982,ikeda1982,chow1983,mallet-paret1986,ikeda1987,mensour1998,adhikari2008,amil2015},
recently some progress was made also for such systems with periodic delay $A\neq0$ theoretically \cite{mueller2018,mueller2019} and experimentally \cite{hart2019,mueller2020,juengling2020,kulminskii2020}.
Eq.~(\ref{eq:DDE}) describes a feedback loop, where a signal is delayed, frequency modulated by a time-varying delay $\tau(t)$, and transformed by a nonlinearity $f$.
The resulting signal is then low-pass filtered with cutoff-frequency $\Theta$ before the next round trip inside the feedback loop.
For $\Theta\gg1$, one obtains the well-studied (singular) limit of large delays \cite{ikeda1982,chow1983,mallet-paret1986,ikeda1987,mensour1998,wolfrum2006,adhikari2008,wolfrum2010,lichtner2011,giacomelli2012,marino2014,amil2015,faggian2018,marino2019},
visible from the time-scale transformation $t^{\prime}=\Theta t$.
In this limit, the low-pass filter approaches the identity and the chaotic dynamics of Eq.~(\ref{eq:DDE}) is governed by the competition of two 1-d maps $x'=f(x)$ and $t'=R(t)$ and their Lyapunov exponents $\lambda_f$ and $\lambda_R$.
Given that $x'=f(x)$ shows chaotic dynamics, $\lambda_f>0$, and the so-called \emph{access map} $t'=R(t)$ is monotonically increasing, Eq.~(\ref{eq:DDE}) basically exhibits two types of chaos \cite{mueller2018,mueller2019}.
By stretching and folding, the map $x'=f(x)$ causes chaotic high-frequency oscillations which are frequency modulated by the time-varying delay.
For $\lambda_R<0$, the access map shows mode-locking
\footnote{In this Letter, the terms mode-locking and quasiperiodic dynamics always refer to the equivalent map on the unit interval, $t'=R(t)\text{ mod } 1$.}.
This means, there is a resonance between the periodic frequency modulation by the delay variation and the round-trip time inside the feedback loop given by the delay itself, leading to low-dimensional \emph{generalized laminar chaos} \cite{mueller2019}.
It is characterized by low-frequency phases with a periodic duration, where the high-frequency oscillations caused by $x'=f(x)$ are mitigated.
For $\lambda_f+\lambda_R<0$, one observes \emph{laminar chaos} \cite{mueller2018}, where the low-frequency phases degenerate to almost constant plateaus, whose levels follow the chaotic dynamics of $x'=f(x)$.
Lacking such resonance, as for constant delay, the access map almost surely shows quasi-periodic dynamics \cite{Note1} with $\lambda_R=0$,
implying that chaotic high-frequency oscillations persist, which characterize high-dimensional \emph{turbulent chaos} \cite{ikeda1980}.

We will show that this distinction has drastic consequences also for delay systems showing chaotic diffusion, which are obtained for nonlinearities such as
\begin{equation}
\label{eq:climbing_sine}
f(x)=x+\mu\sin(2\pi x),
\end{equation}
or, more generally, for unbounded functions $f(x)$ with symmetry properties $f(-x)=-f(x)$ and $f(x+1)=f(x)+1$.
This type of nonlinearity is motivated as in well-known studies of chaotic diffusion in one-dimensional iterated maps $x_{n+1}=f(x_n)$ \cite{geisel1982,schell1982,fujisaka1982},
which capture essential features of driven pendula, Josephson junctions, or phase-locked loops \cite{huberman1980,dhumieres1982}.
Our resulting infinite-dimensional delay-differential equation (DDE), Eq.~(\ref{eq:DDE}) with Eq.~(\ref{eq:climbing_sine}), is an Ikeda-like equation,
which in principle can be realized experimentally by low-pass opto-electronic oscillators \cite{larger2013,chembo2019,hart2019},
by phase-locked loops similar to \cite{wischert1994,schanz2003}, by electronic circuits \cite{zhang2012,juengling2020,karmakar2020}, or by microwave oscillators \cite{dao2013_2}.
As in \cite{ikeda1980}, $x(t)$ can be regarded as a phase variable, which can naturally assume arbitrary large values.
The parameters of the system are $\tau_0$ and $A$ determining the delay variation, the strength $\mu$ of the nonlinearity, and $\Theta$, which sets the overall time-scale.
Note that our system, Eqs.~(\ref{eq:DDE}-\ref{eq:climbing_sine}), for $\mu\rightarrow\infty$, $\Theta\rightarrow0$, s.t., $\mu\Theta=\kappa$ is a constant,
results in $dx(t)/dt=\kappa\sin(2\pi x(R(t)))$, which for constant delay $R(t)=t-1$ is exactly the case studied in \cite{wischert1994,schanz2003,sprott2007}.
The appearance of chaotic diffusion in this system \cite{sprott2007,dao2013_1} suggests that our system, which contains an additional delayed feedback term, is able to generate deterministic diffusion as well.
We will see, firstly, that this is indeed the case, secondly, that fundamentally different mechanisms exist, and thirdly, that the properties of the delay variation in Eq.~(\ref{eq:retarded_argument}) determine the mechanism at work.

\begin{figure}
\includegraphics[width=\linewidth]{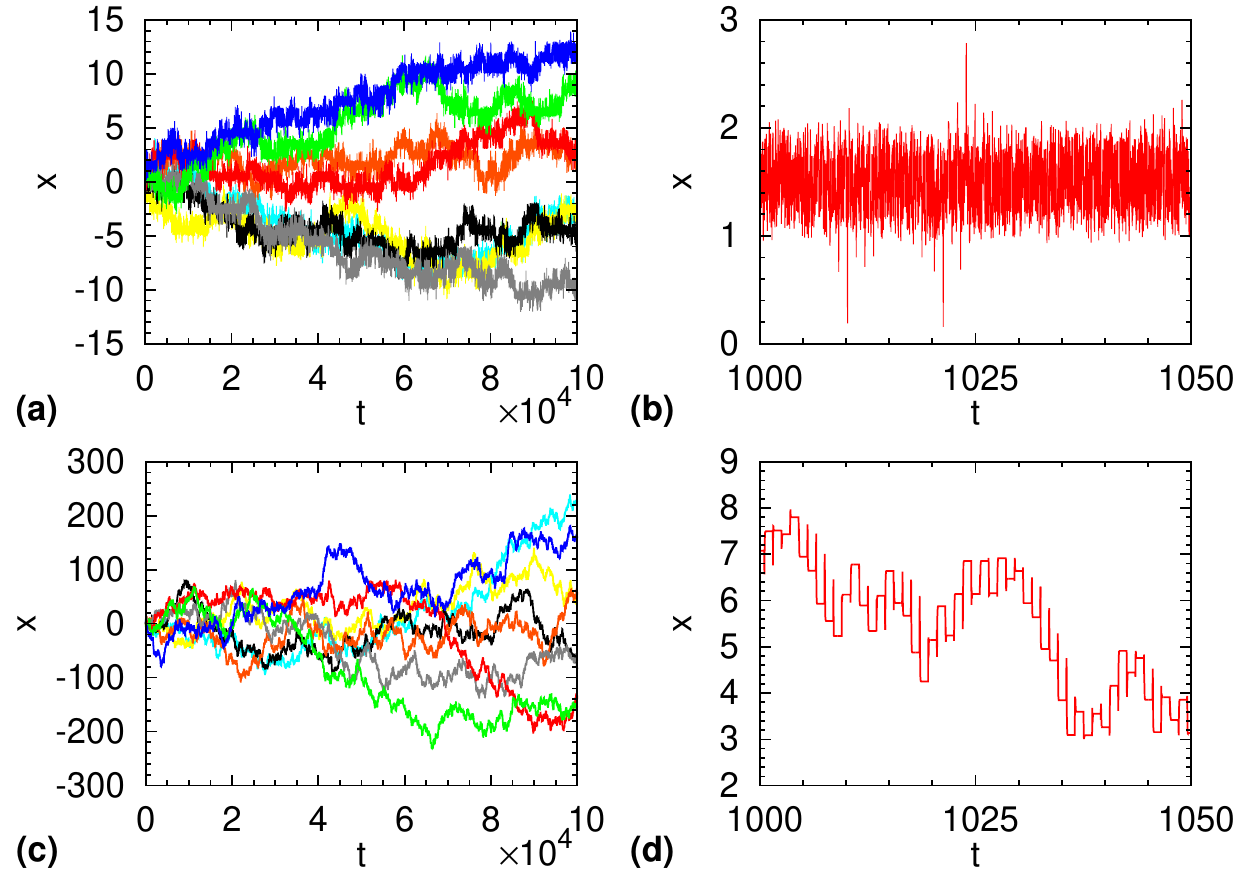}
\caption{\label{fig:trajectories}
Trajectories obtained numerically from Eqs.~(\ref{eq:DDE},\ref{eq:retarded_argument},\ref{eq:climbing_sine}) reflect chaotic diffusion, but with characteristic differences,
when the modulation amplitude of the delay is increased from $A=0$ in (a),(b) to $A=0.9$ in (c),(d), while keeping all other parameters constant ($\Theta=50$, $\tau_0=1$, and $\mu=0.9$).}
\end{figure}

The existence of chaotic diffusion and of different generating mechanisms can be anticipated already from Fig.~\ref{fig:trajectories}.
All trajectories were generated using the same parameters, except for the modulation amplitude, which is zero for the upper, but non-zero for the lower panels.
We see an irregular spreading of the ensemble of trajectories typical for diffusive motion, but the width of the spread for constant delay is by far smaller than for the modulated delay.
In addition, we see characteristic differences of the trajectories on a microscopic time scale, Figs.~\ref{fig:trajectories}(b),(d).
The spreading is quantified by the diffusion constant $D$, which we determined from ensemble-averaged squared-displacements $\left\langle[x(t)-x(0)]^2\right\rangle\simeq Dt$
\footnote{Trajectories of length $10^4\cdot\tau_0$ were generated using the two-stage Lobatto IIIC algorithm with linear interpolation \cite{bellen2003}. Averages were taken over $10^4$ initial functions.}.
For the case $A=0$ of Fig.~\ref{fig:trajectories}, we obtain $D\approx0.0004$, whereas for $A=0.9$, we get $D\approx0.2558$, i.e.,
a modulation of the delay leads to a giant enhancement of the diffusion constant by almost three orders of magnitude.

\begin{figure}
\includegraphics[width=\linewidth]{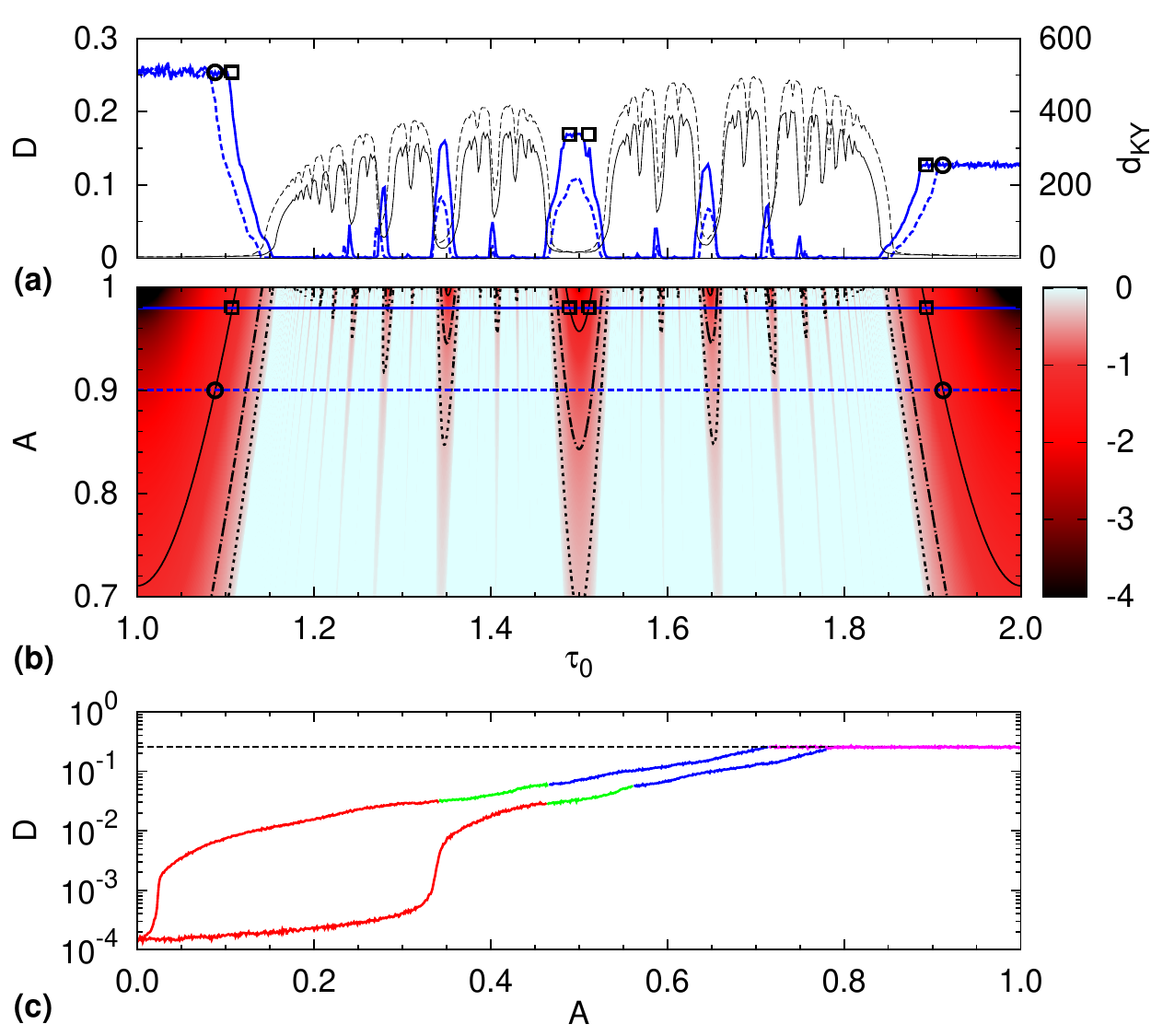}
\caption{\label{fig:arnold_tongues}
(a) A strong, sensitive dependence of the diffusion constant $D$ (thick blue lines) in comparison with the Kaplan-Yorke dimension $d_{KY}$ (thin black lines) on the delay $\tau_0$ is observed ($A=0.98$ (solid lines), $A=0.90$ (dashed lines)),
which can be related to (b) the Arnold tongue structure visible in the Lyapunov chart $\lambda_R(\tau_0,A)$ of the access map.
The full, dashed, and dotted contour lines separate areas of generalized laminar chaos of order $K-1=0,1,2$ (from darkest to brightest), respectively.
Circles and squares mark the boundaries in $\tau_0$ of laminar chaotic diffusion ($K=1$) for $A=0.9$ and $A=0.98$, respectively.
Corresponding marks in (a) are placed at heights predicted by theory.
(c) The variation of $D$ is plotted for two values $\tau_0=1$ and $\tau_0=1.05$ (upper and lower curve, respectively) as the modulation amplitude increases from $A=0$ to $A=1$.
The logarithmic scale shows that the near-zero values (observed also in (a)) are of order $O(10^{-4})$ implying a variation of $D$ of more than three orders of magnitude
while passing through regimes of generalized laminar chaotic diffusion of order $K-1=0,1,2$, and higher (purple, blue, green, red, from right to left).
Laminar chaotic diffusion $K=1$ is special because $D$ hardly varies within the corresponding Arnold tongue area while changing $A$ in (c) or $\tau_0$ in (a) ($\Theta=200$ and $\mu=0.9$).}
\end{figure}

Such a strong variation of $D$ is found in large parts of the accessible $\tau_0$-$A$-plane, $0\leq A<1$, $\tau_0>A/(2\pi)$.
As examples, we plot in Fig.~\ref{fig:arnold_tongues}(a) the diffusion constant $D(\tau_0,A)$ in the interval $1\leq\tau_0\leq2$, for $A=0.98$ and $A=0.90$, respectively.
One finds a strongly structured $\tau_0$-dependence, where $D$ takes large values up to $D\approx0.25$ in the peak regions, but drops to small, near-zero values ($D\approx0.0001$) in certain intervals.
To understand these structures, we plot in Fig.~\ref{fig:arnold_tongues}(b) the corresponding Lyapunov chart $\lambda_R(\tau_0,A)$, i.e., the contour plot visualizing the Lyapunov exponent $\lambda_R$ as function of its parameters $\tau_0$ and $A$.
Comparing the $\tau_0$-dependence of $\lambda_R(\tau_0,0.98)$ and $\lambda_R(\tau_0,0.90)$ with the corresponding variation of the diffusion constants $D(\tau_0,0.98)$ and $D(\tau_0,0.90)$, respectively,
one sees that the $\tau_0$-intervals of near zero $D$ occur when the associated Lyapunov exponents $\lambda_R$ are zero, and the peaks in $D$ occur when $\lambda_R<0$.
In other words, the DDE, Eq.~(\ref{eq:DDE}), shows strong diffusion when the corresponding access map $t^{\prime}=R(t)$ shows mode locking,
whereas the diffusion almost vanishes when the dynamics of the access map is quasi-periodic.
Thus, we established a hitherto unknown connection between the deterministic diffusive dynamics of a DDE, Eq.~(\ref{eq:DDE}), and the Arnold-tongue structure of the map defined via the retarded argument, Eq.~(\ref{eq:retarded_argument}).

\begin{figure}
\includegraphics[width=\linewidth]{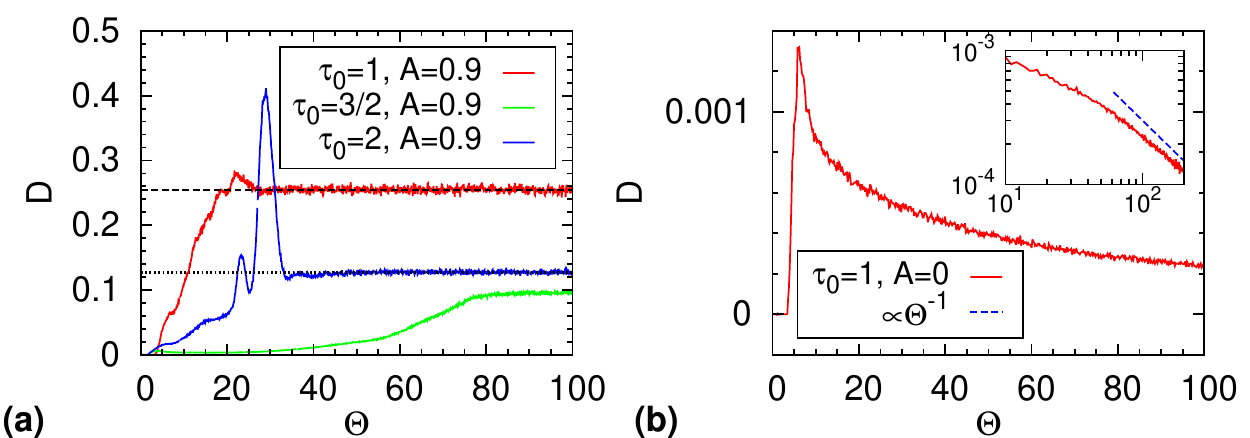}
\caption{\label{fig:D_Theta}
(a) For variable delay and laminar chaotic diffusion ($\tau_0=1$ and $\tau_0=2$) the diffusion constant $D(\Theta\gg1)$ seems to coincide with the expected constant value $D=D_{\mu}/\rho$,
and also for generalized laminar chaos of order one ($\tau_0=3/2$), a constant value is reached.
(b) In contrast, for constant delay, $D$ decays asymptotically as $\Theta^{-1}$ ($\mu=0.9$).}
\end{figure}

A deeper understanding of the strong diffusion part of this phenomenon follows from the observation that in parameter regions with $\lambda_R<0$
the theory of laminar or generalized laminar chaos \cite{mueller2018,mueller2019} applies also to our system.
The most pronounced effect occurs when laminar chaos prevails, i.e., if the condition $\lambda_f+\lambda_R<0$ with $\lambda_f>0$ is fulfilled.
In Fig.~\ref{fig:arnold_tongues}(a), it is fulfilled in the intervals derived from Fig.~\ref{fig:arnold_tongues}(b).
Under this condition, the solution $x(t)$ of Eq.~(\ref{eq:DDE}) consists essentially of a sequence of plateaus, which can start to diffuse if the 1-d map $x^{\prime}=f(x)$ produces chaotic diffusion,
which is the case for Eq.~(\ref{eq:climbing_sine}) if $\mu>\mu_c=0.732...$ \cite{geisel1982}
\footnote{In this parameter regime, the 1-d map can also show anomalous diffusion \cite{korabel2002}.
In an unexpected way, anomalous diffusion can appear also in the corresponding delay system, but not necessarily related to its appearance in the 1-d map, as will be shown in a future publication}.
That this picture applies can be seen from the trajectory of Fig.~\ref{fig:trajectories}(d) with its plateau-like structure.
We expect that the diffusion constant becomes related to that of the Climbing-Sine-Map in the limit $\Theta\gg1$ since then the theory for laminar chaos is valid.
Fig.~\ref{fig:D_Theta}(a) confirms this expectation: for $\Theta\geq25$, the diffusion constant already takes its asymptotic, non-zero value, which near $\tau_0=1$ is directly $D_{\mu}$, the one given by the Climbing-Sine-Map.

For a quantitative understanding of the structures in Fig.~\ref{fig:arnold_tongues}(a), we recall that the solution of DDEs can be generated iteratively by the method of steps \cite{bellman1965},
where solution segments $x_n(t)$ with $t\in(t_{n-1},t_n]$ and $t_{n-1}=R(t_n)$ are generated from the preceding segment $x_{n-1}(t)$ starting with an initial function $x_0(t)$ with $t\in(t_{-1},t_0]$.
For Eq.~(\ref{eq:DDE}), in the singular limit $\Theta=\infty$, this function mapping can be written as
\begin{equation}
\label{eq:limit_map}
x_{n+1}(t)=f(x_n(R(t))),
\end{equation}
where the low-pass filter of the feedback loop is neglected.
Altough one usually observes the diffusive behavior of the scalar $x(t)$ in continuous time, it is rather the function $x_n$, an infinite-dimensional object, which diffuses in discrete time $n$.
For our purposes, it is sufficient to observe the dynamics of one 'component' of $x_n$, e.g., $x_n(t)$ at the endpoint $t=t_n$.
If the condition for laminar chaos is fulfilled, $R$ shows mode-locking with a rational rotation number $\rho=p/q$, which is the average number of delay periods covered per round-trip inside the feedback loop.
Then $x_n$ consists of $p$ plateaus whose levels are mapped forward by $f$ from Eq.~(\ref{eq:climbing_sine}), i.e., $x_{n+1,i}=f(x_{n,i})$, where $x_{n,i}$ denotes the $i$th plateau of $x_n$.
Therefore, if iterations $x^{\prime}=f(x)$ display chaotic diffusion with diffusion constant $D_{\mu}$, the plateaus show the same chaotic diffusion in discrete time $n$.
To obtain the diffusion constant $D$ for Eq.~(\ref{eq:DDE}) in continuous time $t$,
we observe that the average length $\langle t_n-t_{n-1} \rangle$ of the solution segments is given by the rotation number $\rho$ resulting in $D=D_{\mu}/\rho$.
This can be observed in Fig.~\ref{fig:arnold_tongues}(a): in the left $\tau_0$-interval near $\tau_0=1$, we have mode-locking with $\rho=1$ and therefore $D=D_{\mu}\approx0.254$,
whereas near $\tau_0=2$, we have $\rho=2$ with $D=D_{\mu}/2$, and in the middle region near $\tau_0=3/2$, we get $D=2D_{\mu}/3$ because $R$ shows mode-locking with $\rho=3/2$.
The picture of independently diffusing plateaus is strictly true only in the limit $\Theta\rightarrow\infty$.
For large but finite $\Theta$, the plateaus are coupled through the finite relaxation rate of the low-pass filter, but this does not affect the diffusion constant $D$,
as shown in Fig.~\ref{fig:D_Theta}(a) for cases with one ($\rho=1$) or two ($\rho=2$) plateaus per state-interval.
This behavior implies that for \textit{laminar chaotic diffusion}, the Kaplan-Yorke dimension of the DDE dynamics (considering $x(t)\text{ mod }1$) is relatively small (see Fig.~\ref{fig:arnold_tongues}(a)
\footnote{To determine the Kaplan-Yorke dimension, we adapted the method from \cite{farmer1982}, where the linearized DDE was discretized with step size $\tau_{\text{max}}/M$ with $\tau_{\text{max}}=\tau_0+A/(2\pi)$ and $M=2000$}),
as in non-diffusive cases \cite{mueller2018}.

In general, in the mode-locked region of the access map, i.e., for $\lambda_R<0$, one can always find a finite integer $K$ such that $\lambda_f+k\lambda_R<0$ for $k\geq K$.
Given that $\lambda_f>0$, this is, for $K\geq2$, the condition for generalized laminar chaos of order $K-1$ \cite{mueller2019}, thus generalizing the above laminar chaotic case $K=1$.
For $K\geq2$, the role of the plateaus seen for $K=1$, is taken by polynomials of order $K-1$, and the effective dimensions are of order $O(K)$, i.e., still small if $K$ is small.
Also for generalized laminar chaos, where we currently have no good quantitative theory for the diffusion constants,
we find numerically that for $\Theta\rightarrow\infty$, the diffusion constant converges to finite, non-zero values, as seen in Fig.~\ref{fig:D_Theta}(a) for generalized laminar chaos of order one.
These finite values can also be inferred from Fig.~\ref{fig:arnold_tongues}(a) by changing $\tau_0$ from the center of an Arnold tongue, where laminar chaotic diffusion is prevailing,
towards its boundary, where, due to decreasing $\vert\lambda_R\vert$, one passes through all orders of \textit{generalized laminar chaotic diffusion}.
This is better seen for fixed $\tau_0$ by varying the modulation amplitude as in Fig.~\ref{fig:arnold_tongues}(c), i.e., by traversing the primary Arnold tongue vertically.
Probably, this is a well-suited scenario for experimentally observing the giant enhancement of the diffusion constant.

These scenarios for $\lambda_R<0$ should be contrasted with the other extreme of near-zero diffusion,
which in Fig.~\ref{fig:arnold_tongues}(b) occurs when the Lyapunov exponent of the access map, Eq.~(\ref{eq:retarded_argument}), vanishes, $\lambda_R=0$.
In Fig.~\ref{fig:D_Theta}(b), we show the corresponding typical $\Theta$-dependence of the diffusion constant $D$.
We observe that, in contrast to (generalized) laminar chaotic diffusion, $D$ decays to zero as $D\sim\Theta^{-1}$ for large $\Theta$.
To understand this law, we first observe that $\lambda_R=0$ means that the dynamics of the access map $t^{\prime}=R(t)$ is (almost surely) quasi-periodic.
Therefore, the DDE of Eq.~(\ref{eq:DDE}) can be transformed to constant delay \cite{otto2017,mueller2017}.
So it is sufficient to understand the origin of chaotic diffusion from Eq.~(\ref{eq:DDE}) for constant delay $R(t)=t-\tau_0$ with $\tau_0=1$.

We know that due to the low-pass filter with cutoff-frequency $\Theta$, solutions $x(t)$ of Eq.~(\ref{eq:DDE}) show (random) oscillations on a time-scale $1/\Theta$.
This suggests that the term $\sin(2\pi x(t-1))$ can be replaced by a noise term $\xi_{\Theta}(t)$ with zero mean and exponentially decaying correlations with correlation length $1/\Theta$,
i.e., $\left\langle\xi_{\Theta}(t)\xi_{\Theta}(t^{\prime})\right\rangle=1/2\exp(-\Theta|t-t^{\prime}|)$.
This approximation turns our DDE into a stochastic DDE of the form
\begin{equation}
\label{eq:SDDE}
\frac{dy(t)}{dt}=-\Theta y(t)+\Theta y(t-1)+\mu\Theta\xi_{\Theta}(t),
\end{equation}
where the approximate solution is denoted as $y(t)$.
Assuming Gaussianity of $\xi_{\Theta}(t)$, Eq.~(\ref{eq:SDDE}) can be considered as that of an Ornstein-Uhlenbeck process (with colored noise) extended by the delay term $\Theta y(t-1)$,
or alternatively as delayed feedback control $\Theta(y(t-1)-y(t))$ \cite{pyragas1992}, which is applied to a (generalized) Wiener process \cite{galleani2006}.
Such systems have been studied previously for white noise \cite{gushchin1999,ando2017} but also for colored noise \cite{budini2004}.
According to the general theory of linear equations, such as Eq.~(\ref{eq:SDDE}), the behavior of the mean-squared displacement (assuming $\left\langle y(0)\right\rangle=\left\langle y(t)\right\rangle=0$) can be expressed as \cite{budini2004}
$\left\langle y^2(t)\right\rangle=(\mu\Theta)^2\int_0^t\int_0^ty_0(t-t^{\prime})\left\langle\xi_{\Theta}(t^{\prime})\xi_{\Theta}(t^{\prime\prime})\right\rangle y_0(t-t^{\prime\prime})\,dt^{\prime}\,dt^{\prime\prime}$,
where $y_0(t)$ is the Green's function or fundamental solution of the deterministic part of Eq.~(\ref{eq:SDDE}).
For large values of $t$, the latter becomes constant $\lim_{t\rightarrow\infty}y_0(t)=1/(1+\Theta)$ \cite{gushchin1999}, so that for large $t$, we have
$\left\langle y^2(t)\right\rangle\sim(\mu\Theta)^2/(1+\Theta)^2\int_0^t\int_0^t\left\langle\xi_{\Theta}(t^{\prime})\xi_{\Theta}(t^{\prime\prime})\right\rangle\,dt^{\prime}\,dt^{\prime\prime}$,
and in the same limit the double integral evaluates to $t/\Theta$ resulting in $\left\langle y^2(t)\right\rangle\sim Dt$ with  $D=\mu^2\Theta/(1+\Theta)^2$.
The same result is obtained if in Eq.~(\ref{eq:SDDE}) one replaces $\Theta^{1/2}\xi_{\Theta}(t)$ by its large-$\Theta$ limit,
which is Gaussian white noise $\xi(t)$ with $\left\langle\xi(t^{\prime})\xi(t^{\prime\prime})\right\rangle=\delta(t^{\prime}-t^{\prime\prime})$, as treated in \cite{gushchin1999,ando2017}.

This confirms the numerically observed behavior of the diffusion constant for \textit{turbulent chaotic diffusion} shown in Fig.~\ref{fig:D_Theta}(b)
\footnote{While our analytical considerations are able to reproduce the correct asymptotic $\Theta$-dependence of the diffusion coefficient,
the assumption of exponentially decaying noise correlations is too simple to reproduce the correct, non-trivial $\mu$-dependence of the diffusion constant, which will be treated in a separate publication.}.
This is a counterintuitive result because diffusion becomes progressively weaker the more turbulent the system behaves as measured by its effective dimensions (see Fig.~\ref{fig:arnold_tongues}(a)).
Note also that for constant delay individual components $x_n(t)$ of the solution segments show vanishing diffusion for $\Theta\rightarrow\infty$,
whereas for $\Theta=\infty$, they diffuse chaotically with $D=D_{\mu}$ (see Eq.~(\ref{eq:limit_map})).

The $\Theta^{-1}$-law implies that a delay modulation can, in principle, cause an increase in the diffusion constant by arbitrary many orders of magnitude.
The only limitations are the experimentally accessible $\Theta$-regime and the influence of additional noise from the experimental environment.

In summary, we found in a typical class of delay systems new mechanisms for deterministic chaotic diffusion with vastly different diffusion constants, namely turbulent and laminar chaotic diffusion,
which are obtained by changing, e.g., the modulation amplitude of the delay.
Since our main results are very general, because they do not depend on the detailed form of the periodic delay modulation, or of the periodic part of the delayed nonlinearity,
we expect that they can be found experimentally in especially designed electro-optical devices or in purely electronic realizations, but presumably also in a much wider class of systems
including, e.g., Josephson junctions if a modulated delay can be introduced.

\begin{acknowledgments}
We thank the anonymous referees for valuable suggestions.
The authors gratefully acknowledge funding by the Deutsche Forschungsgemeinschaft (DFG, German Research Foundation) - 438881351; 456546951.
\end{acknowledgments}

\bibliography{references}

\end{document}